\documentclass[12pt,preprint,psfig]{aastex}

\usepackage{emulateapj5}
\usepackage{graphicx}

\def\ltsima{$\; \buildrel < \over \sim \;$}

\def\lsim{\lower.5ex\hbox{\ltsima}}
\def\gtsima{$\; \buildrel > \over \sim \;$}
\def\gsim{\lower.5ex\hbox{\gtsima}}

\begin{document}
\title{Prospects for Redshifted 21-cm observations of quasar HII regions}

\author{J. Stuart B. Wyithe\altaffilmark{1}, Abraham Loeb\altaffilmark{2},
David G. Barnes\altaffilmark{1}}

\email{swyithe@physics.unimelb.edu.au; aloeb@cfa.harvard.edu; dbarnes@physics.unimelb.edu.au}

\altaffiltext{1}{University of Melbourne, Parkville, Victoria, Australia}

\altaffiltext{2}{Astronomy department, Harvard University, 60 Garden
St., Cambridge, MA 02138}

\begin{abstract}
\noindent 
The introduction of low-frequency radio arrays over the coming decade
is expected to revolutionize the study of the reionization
epoch. Observation of the contrast in redshifted 21cm emission between
a large \ion{H}{2} region and the surrounding neutral IGM will be the
simplest and most easily interpreted signature. We assess the
sensitivity of three generations of planned low-frequency arrays to
quasar-generated
\ion{H}{2} regions. We find that an instrument like the planned {\it
Mileura Widefield
Array Low-Frequency Demonstrator} (LFD) will be able to obtain good signal
to noise on \ion{H}{2} regions around the most luminous quasars, and
determine some gross geometric properties, e.g. whether the \ion{H}{2}
region is spherical or conical. A hypothetical follow-up instrument with 10 times the collecting area of the LFD (MWA-5000) will be capable of mapping the detailed geometry of
\ion{H}{2} regions, while SKA will be capable of detecting very
narrow spectral features as well as the sharpness of the
\ion{H}{2} region boundary. The SKA will most likely be limited by
irreducible noise from fluctuations in the IGM itself. The MWA-5000 (and
even the LFD under favorable circumstances) will discover
serendipitous \ion{H}{2} regions in widefield observations. We
estimate the number of \ion{H}{2} regions which are expected to be
generated by quasars based on the observed number counts of quasars at
$z\sim 6$. Assuming a late reionization at $z\sim6$ we find that there
should be several tens of quasar \ion{H}{2} regions larger than 4Mpc
at $z\sim6-8$ per field of view. Identification of \ion{H}{2} regions
in forthcoming 21cm surveys can guide a search for bright galaxies in
the middle of these regions.  Most of the discovered galaxies would be
the massive hosts of dormant quasars that left behind fossil
\ion{H}{2} cavities that persisted long after the quasar emission
ended, owing to the long recombination time of intergalactic
hydrogen. A snap-shot survey of candidate \ion{H}{2} regions selected
in redshifted 21cm image cubes may prove to be the most efficient
method for finding very high redshift quasars and galaxies.

\end{abstract}

\keywords{cosmology: theory - galaxies: formation}

\section{Introduction}

The Gunn-Peterson (1965, hereafter GP) troughs in the spectra of the
most distant quasars at redshifts $z\sim 6.3$--$6.4$ (Djorgovski et
al.~2001; Becker et al. 2001; Fan et al.~2004) hint that the
reionization of cosmic hydrogen was completed only a billion years
after the big bang.  Unfortunately, the troughs only set a lower limit
of $\sim 10^{-3}$ on the volume averaged neutral fraction (Fan et
al.~2001; White et al. 2003). This leaves open the question of whether
reionization was completed only at $z\sim6$ so that the GP trough is
due to a significantly neutral IGM, or whether it is only a residual
neutral fraction left over from reionization at an earlier epoch that
is responsible for the observed Ly$\alpha$ absorption. We can start to
answer this question by noting that the sizes of \ion{H}{2} regions
generated by luminous quasars (Madau \& Rees~2000; Cen \& Haiman~2000;
Barkana \& Loeb 2003) at $z>6$ imply a neutral fraction that is of
order unity at that time (Wyithe \& Loeb~2004a; Wyithe, Loeb \&
Carilli~2005). In addition Messinger \& Haiman (2004) analyzed
SDSS~1030+0524 and found that the shape of the absorption spectrum
near the edge of the \ion{H}{2} region requires a large neutral
fraction. These findings appear in conflict with evidence presented by
the WMAP satellite in favor of early reionization (Kogut et al.~2003;
Spergel et al.~2003). Indeed, if reionization ended only at $z\sim6$,
then more extended reionization scenarios involving a massive early
population of stars may be required (Wyithe \& Loeb~2003a; Cen~2003;
Furlanetto \& Loeb 2005).

The large optical depth for Ly$\alpha$ absorption in a neutral
intergalactic medium (IGM) limits its effectiveness in probing reionization
at redshifts beyond $z\sim6$. 
A better probe of the reionization history involves the redshifted 21cm
emission from neutral hydrogen in the IGM. Various experiments are planned
to measure this emission and are summarized in \S\ref{21cmtel}.  Several
probes of the reionization epoch in redshifted 21cm emission have been
suggested. These include observation of the emission as a function of
redshift averaged over a large area of the sky. This provides a direct
probe of the evolution in the neutral fraction of the IGM, the so-called
global step (Shaver, Windhorst, Madau \& de~Bruyn~1999; Gnedin \&
Shaver~2004). A more powerful probe will be provided by observation of the
power-spectrum of fluctuations together with its evolution with redshift.
This observation would trace the evolution of neutral gas with redshift as
well as the topology of the reionization process (e.g. Tozzi, Madau, Meiksin \&
Rees~2000; Furlanetto, Hernquist \& Zaldarriaga~2004; Loeb \& Zaldarriaga
2004; Barkana \& Loeb 2005a,b,c). Finally observation of individual
\ion{H}{2} regions will probe quasar physics as well as the evolution of
the neutral gas (Wyithe \& Loeb~2004b; Kohler, Gnedin, Miralda-Escude \&
Shaver~2005). Kohler et al.~(2005) have generated synthetic spectra using
cosmological simulations. They conclude that quasar \ion{H}{2} regions will
provide the most prominent individual cosmological signals.  In this paper we
concentrate on the signature of individual quasar \ion{H}{2} regions with
emphasis on the capabilities of planned telescopes.

While most discussion of reionization signatures in redshifted 21cm
radiation has centered on indirect probes of the earliest galaxies
through their statistical effect on the IGM, the detection of \ion{H}{2} regions
will provide specific sites where high redshift galaxies and quasars must be
present.  The absence of neutral hydrogen within \ion{H}{2} regions
means that these are also sites where the sources can be studied with less
interference from rest-frame Ly$\alpha$ absorption. Indeed for this
reason Cen~(2003) has suggested \ion{H}{2} regions around known high
redshift quasars as sites to search for high-redshift
galaxies. While known high redshift galaxies are rare, the advent of
low-frequency radio telescopes will enable discovery of a large number of
\ion{H}{2} regions. Some of these regions will house luminous quasars,
while most will house galaxies and groups of galaxies. Targeted
near to mid-IR observations of 21cm selected \ion{H}{2} regions will help define the
role of different classes of sources in the reionization of the IGM.

A radio telescope is limited in sensitivity by two separate factors. The first
corresponds to its physical configuration, such as the limit on
resolution (set by the largest baseline) or the flux limit that can be
detected over a reasonable observing time at the required resolution
(set by the collecting area). Second, the theoretical limits based on
the telescope dimensions may not be realized due to difficulties in
achieving a perfect calibration, or due to problems introduced by
foregrounds and the ionosphere. There has been significant discussion
in the literature regarding foregrounds. While the fluctuations due to
foregrounds are orders of magnitude larger than the reionization
signal (e.g. di~Matteo, Perna, Abel \& Rees~2002; Oh \& Mack~2003),
the foreground spectra are anticipated to be smooth. Since the
reionization signatures include fluctuations in frequency as well as
angle, they can be removed through continuum subtraction (e.g. Gnedin
\& Shaver~2004; Wang, Tegmark, Santos \& Knox~2005) or removed using
the differences in symmetry from power-spectra analyses (Morales \&
Hewitt~2004; Zaldarriaga et al.~2004). In this paper we concentrate on
the theoretical limitations of planned 21cm instruments. The practical
difficulties introduced by calibration and foreground removal
etc. need to be addressed by full simulations of interferometric
observations, and ultimately with early data sets from new facilities.

The paper is organized as follows. We begin in \S\ref{21cmtel} by
summarizing the design properties of low frequency telescopes that are
currently being planned or constructed. We then discuss the response
of a synthesis array to the 21cm signature of an \ion{H}{2} region in
the presence of fluctuating background emission from density
inhomogeneities in the IGM (\S\ref{response}). The sizes of known
quasar \ion{H}{2} regions, and our models for quasar \ion{H}{2}
regions are described in \S\ref{properties} and \S\ref{model}. Mock
observations are then presented in \S\ref{observations} to enable
assessment of the performance that is possible from the planned
instruments. In \S\ref{counts} we estimate the number of \ion{H}{2}
regions expected per field of view. We then discuss some science goals
in \S\ref{science} before summarizing our conclusions in
\S\ref{conclusion}.

\section{Planned low frequency telescopes}
\label{21cmtel}

A variety of instruments capable of detecting emission from neutral
gas in the high redshift IGM are planned or are under construction. Interferometric arrays currently under construction include
the Primeval Structure
Telescope\footnote{http://web.phys.cmu.edu/~past/} (PAST), the Low
Frequency Array \footnote{http://www.lofar.org/} (LOFAR) and a
re-fitting of the VLA with low-frequency (200MHz) receivers (L. Greenhill et al). 
Of these, the low frequency VLA will have the smallest
collecting area. However this system has been especially designed with
observation of
\ion{H}{2} regions around known high redshift quasars in mind. 
Conversely the PAST experiment is designed to perform a deep
observation of the north Celestial Pole with an aim of detecting 
the power-spectrum of 21cm fluctuations. In its initial form PAST will
not have a pointable primary beam and so is less suited to observing
known \ion{H}{2} regions. The LOFAR telescope will consist of phased
arrays of cross-dipoles grouped in a large number of stations or
tiles. Each tile has a large primary beam covering many square
degrees. LOFAR will be capable of operating with pointed observations
over a wide band-pass at redshifts from 5-20. While the northern
European site on which LOFAR is being built is not an ideal radio
environment, LOFAR will be complementary to the planned Mileura
Widefield Array\footnote{http://web.haystack.mit.edu/arrays/MWA/}
which we describe below.

The Mileura Wide-field Array (MWA) is a collaboration of the
Massachusetts Institute of Technology, Harvard--Smithsonian Center for
Astrophysics, an Australian University Consortium led by Melbourne and
the Australian National Universities, the Australia Telescope National
Facility, and the Western Australian government.  This paper is
concerned only with observations using the low-frequency portion of
the planned MWA (the MWA-Low Frequency Demonstrator) which we simply
refer to as the LFD. The telescope will be located at the Mileura
radio quiet site in Western Australia, and the low frequency portion
of the array would consist of 500 tiles each containing 16 cross
dipoles which operate over the range 80-300MHz. The tiles would be
spread to cover baselines of up to $\sim2$km. In collecting area the
LFD would be comparable to the VLA. However the correlator/receiver
system will be able to handle a much larger number of channels than
will be available on the low frequency VLA.  A hypothetical follow up
to the LFD would comprise 10 times the collecting area of the LFD,
with baselines of $\sim10$s to $100$s of kilometers (we refer to this
as MWA-5000). The physical attributes of the MWA-5000 would be
comparable to LOFAR as an instrument for study of
reionization. Finally a Square Kilometer Array (SKA) would become the
next generation radio telescope. An SKA would have the collecting area
of $\sim10$ times the MWA-5000, with baselines up to $\sim1000$km or
greater. In this paper we concentrate our discussion on the
capabilities of the LFD, MWA-5000 and SKA. However discussion of the
LFD is also broadly applicable to the low-frequency VLA, while the
discussion of the MWA-5000 could be applied to LOFAR.

\section{Telescope Response}
\label{response}

\begin{figure*}[t]
\epsscale{2.} \plotone{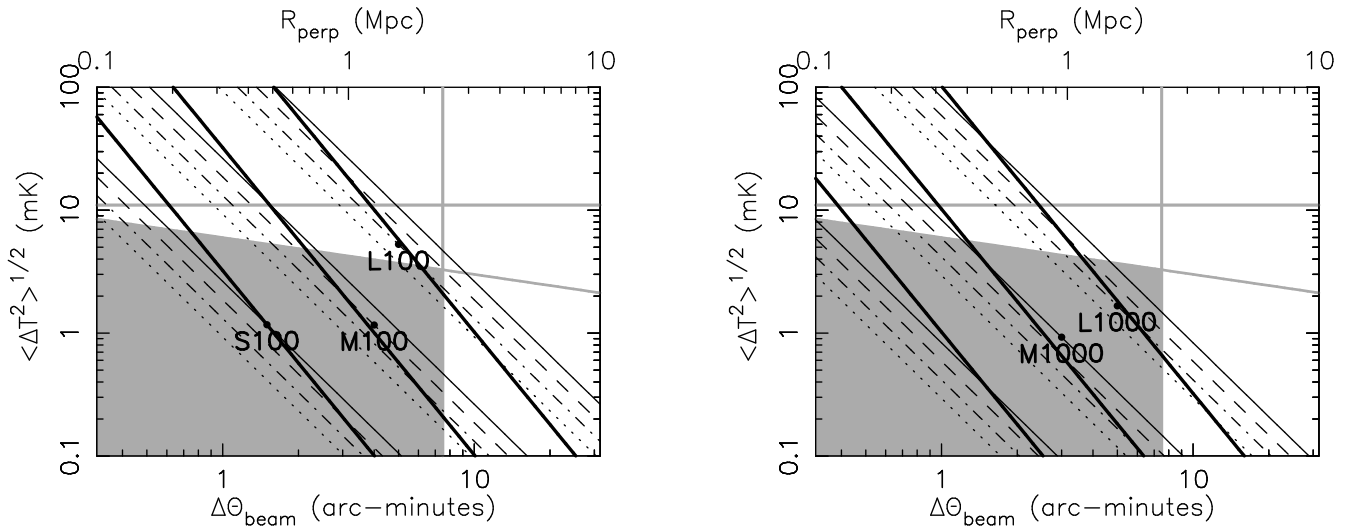}
\caption{\label{fig0} Response of different low-frequency telescopes relative 
to the expected signal of \ion{H}{2} regions. The shaded region is
bounded from the right by half of the size of a 5Mpc HII region, and
from above by the variance in brightness temperature due to IGM
density fluctuations smoothed on the scale shown and averaged in a
0.5MHz band pass. The horizontal grey line at 11mk corresponds to half
of the 22mK maximum contrast between an \ion{H}{2} region and the
surrounding neutral IGM at $z=6.5$. The telescope sensitivity should
lie within the shaded region. The
sets of dark lines show the rms noise evaluated from the radiometer
equation as a function of Gaussian beam size for 0.5, 1, 2 and 4MHz
bins (right to left), and for a frequency bin size that matches the physical scale of
the beam (thick line). The three sets of lines correspond to a
collecting area of the LFD, MWA-5000 and SKA (right to left). The left and
right panels are for a 100hr and a 1000hr integration
respectively. The labels L100 and L1000 show the points corresponding to the LFD
with 100 and 1000 hour integrations respectively
(Figure~\ref{fig3}), while the labels M100 and M1000 show the points
corresponding to the MWA-5000 with 100 and 1000 hour integrations
(Figures~\ref{fig3} and ~\ref{fig4}). The label S100
corresponds to the SKA with a 100 hour integration
(Figure~\ref{fig6}). }
\end{figure*}

In this section we discuss the response of a phased array to the
brightness temperature contrast introduced by an \ion{H}{2}
region. Assuming that calibration can be performed ideally, and that
foreground subtraction is perfect, the root-mean-square fluctuations in
brightness temperature are given by the radiometer equation
\begin{equation}
\langle\Delta T^2\rangle^{1/2} = \frac{C\lambda^2T_{\rm sys}}{A_{\rm
tot}\Omega_{\rm b}\sqrt{t_{\rm int}\Delta\nu}},
\end{equation}
where $\lambda$ is the wavelength, $T_{\rm sys}$ is the system
temperature, $A_{\rm tot}$ the collecting area, $\Omega_{\rm b}$ the
effective solid angle of the synthesized beam in radians, $t_{\rm
int}$ is the integration time, $\Delta\nu$ is the size of the
frequency bin, and $C$ is a constant that describes the overall
efficiency of the telescope. We optimistically adopt $C=1$ in this
paper. In units relevant for upcoming telescopes, and at $\nu=200$MHz
we find
\begin{eqnarray}
\label{noiseeqn}
\nonumber
\Delta T &=& 7.5\left(\frac{1.97}{C_{\rm beam}}\right)\mbox{mK} \left(\frac{A}{A_{\rm LFD}}\right)^{-1}\\
&\times&\left(\frac{\Delta\nu}{1\mbox{MHz}}\right)^{-1/2}\left(\frac{t_{\rm int}}{100\mbox{hr}}\right)^{-1/2}\left(\frac{\Delta\theta_{\rm beam}}{5^\prime}\right)^{-2}.
\end{eqnarray}
Here $A_{\rm LFD}$ is the collecting area of a phased array consisting
of 500 tiles each with 16 cross-dipoles [the effective collecting
area of an LFD tile with $4\times4$ cross-dipole array with 1.07m
spacing is $\sim17-19$m$^2$ between 100 and 200MHz (B. Correy, private
communication)]. The system temperature at 200MHz will be dominated by
the sky and has a value $T_{\rm sys}\sim250$K. The frequency range
over which the signal is smoothed is $\Delta\nu$ and
$\Delta\theta_{\rm beam}$ is the size of the synthesized beam in
arc-minutes. The beam size $\Delta\theta_{\rm beam}$ can be
interpreted as the radius of a hypothetical top-hat beam, or as the
variance of a hypothetical Gaussian beam. The corresponding values of
the constant $C_{\rm beam}$ are 1.97 and 1 respectively.

Noise from the telescope is not the only form of confusion. The
density fluctuations in the IGM also produce a form of random noise
relative to the detection of \ion{H}{2} regions. Of course these
fluctuations do not reduce with observing time. At times where the
neutral fraction is near unity, and on large enough scales the
power-spectrum of emission from density fluctuations may be calculated
directly from the linear power-spectrum of primordial
fluctuations. The power-spectrum is spherically symmetric in space,
and may be approximated by a power-law with $\langle\Delta
T^2\rangle^{1/2}\propto\theta_{\rm beam}^{-1/3}$ (Furlanetto \& Briggs~2004). 
The normalization depends on the details of the frequency smoothing scale
and on the neutral fraction etc. We take the fiducial value of $3$mK at 10$^\prime$
for a smoothing of 5MHz (Furlanetto \& Briggs~2004). The resulting
power-spectrum is shown in Figure~\ref{fig0}. While telescope noise is
spread across the entire image cube, fluctuations from the cosmic web are
present only outside the ionized regions. If the neutral fraction is near
unity, then the contrast of the \ion{H}{2} region will be much larger than
the variance due to the cosmic web. Therefore if the goal is to measure the
properties of the \ion{H}{2} region, then the benefit of increasing the
integration time becomes small once the noise reaches the level of the
fluctuations in the IGM emission. Optimal use of the telescope for observing
\ion{H}{2} regions therefore calls for integration times and resolutions
that result in instrumental noise that is near or just below the level
of variance due to the cosmic web.

\begin{table*}[t]
\begin{center}
\caption{\label{tab1} Properties of $z>6$ quasars. Radii $R$ are quoted in Mpc.}
\begin{tabular}{cccccc}
\hline 
Source & $M_{1450}$ & $z$ & $z_{GP}$ & $z_{host} - z_{GP}$ & $\bar{R}\pm\sigma_R$ \\
\hline 

J1148+5251 & -27.82$^a$ & $6.419\pm0.001^b$ & $6.325\pm0.02^c$ & $0.09\pm0.02$ & 4.89$\pm$1.09 \\ 

J1030+0524 & -27.15$^d$ & $6.311\pm0.005^g$ & $6.178\pm0.005^c$ & $0.133\pm0.007$ & 7.50$\pm$0.39  \\

J1048+4637 & -27.55$^a$ & $6.203\pm0.005^g$ & $6.16\pm0.03^a$ & $0.04\pm0.03$ &   2.34$\pm$1.76  \\

J1623+3112 & -26.67$^e$ & $6.22\pm0.03^e$ & $6.16\pm0.03^e$ & $0.08\pm0.04$ & 4.65$\pm$1.74  \\

J1602+4228 & -26.82$^e$ & $6.07\pm0.03^e$ & $5.95\pm0.03^e$ & $0.12\pm0.04$ &   7.36$\pm$2.45  \\ 

J1630+4012 & -26.11$^a$ & $6.065\pm0.005^g$ & $5.98\pm0.03^a$ & $0.085\pm0.03$ &   5.22$\pm$1.84  \\

J1306+0356 & -27.19$^d$ & $5.99\pm0.02^f$ & $5.90\pm0.03^d$ & $0.09\pm0.04$ &   5.68$\pm$2.52  \\ \hline

\multicolumn{6}{c}{$^a$Fan et al. (2003); $^b$Walter et al. (2003); $^b$Bertoldi et al. (2003); $^c$White et al. (2003)} \\

\multicolumn{6}{c}{$^d$Fan et al. (2001); $^e$Fan et al. (2004); $^f$Maiolino et al. (2003); $^g$Iwamuro et al. (2004)}

\end{tabular}
\end{center}
\end{table*}

For a given telescope and synthesized beam size, the minimum
integration time necessary for the observation of an \ion{H}{2} region
is therefore set by the requirement that the noise be smaller than both half the
maximum contrast, which is 11mK at $z=6.5$ (plotted as the horizontal
line in Figure~\ref{fig0}), and the fluctuations in the IGM
surrounding the \ion{H}{2} region. In addition the maximum beam size is set
by the requirement that the resolution be smaller than the feature
being observed. A spherical \ion{H}{2} region at $z\sim6.5$ of radius
5Mpc subtends an angular radius of $\sim15^\prime$ (half of which is
plotted as the vertical line in Figure~\ref{fig0}). 
The shaded region in Figure~\ref{fig0} shows the region bounded from above by
variance in emission from the IGM, and from the right by resolution requirements. The
parameters describing the observation of an \ion{H}{2} region should
place the noise inside, but near the top of this shaded area.

For comparison with these requirements we overplot the sensitivity of
different telescopes in Figure~\ref{fig0}. The ideal root-mean-square
noise is plotted (sets of curves from right to left) for the LFD, for
the MWA-5000 (with $A_{\rm MWA-5000}=10A_{\rm LFD}$), and for the SKA
(with $A_{\rm SKA}=100A_{\rm LFD}$). The left and right panels
represent 100 and 1000 hour integrations respectively. Each set of
curves shows the instrumental noise corresponding to 4, 2, 1 and
0.5MHz smoothing (thin lines left to right). In each case the thick
lines in Figure~\ref{fig0} show the noise as a function of synthesized
beam size where the scale of the frequency smoothing is chosen to
match the physical scale of the beam at the redshift of the
signal. The noise behaves as $(\Delta\theta_{\rm beam})^{-3}$ rather
than $(\Delta\theta_{\rm beam})^{-2}$ in this case.

The telescopes described in the previous section therefore meet the
requirements for observation of \ion{H}{2} regions over long
integrations, as may be seen by inspecting Figure~\ref{fig0}. Of
course Figure~\ref{fig0} describes the most optimistic case, as the
contrast will be lower than 22mK if the IGM is partially
ionized. Moreover the integration time should be sufficiently long
that the contrast is measured at a signal-to-noise significantly in
excess of unity. Furthermore, for a spherical \ion{H}{2} region the
beam-size should be much smaller than the angular extent of the
\ion{H}{2} region otherwise the contrast is reduced due to dilution of
the signal. Similarly, if the spectral resolution corresponds to a
size in the Hubble flow that is not smaller than the extent of the
\ion{H}{2} region, then the contrast will be reduced due to dilution
of the signal during smoothing in frequency. Note that it is {\em
cheaper} in terms of telescope time to gain resolution along the
line-of-sight than perpendicular to the line-of-sight.

The contrasts in Figure~\ref{fig0} and the remainder of this paper
assume a neutral fraction of unity. However to relate the results in
this paper to a neutral fraction that is smaller than unity, the
integration times discussed in this paper should be increased by a
factor of $x_{\rm HI}^{-2}$. In addition, as the neutral fraction
drops towards the end of reionization, the variance of IGM
fluctuations can rise due to the appearance of ionized regions. This
variance will swamp the signal of individual \ion{H}{2} regions. Our
simulations therefore correspond to the IGM before it has been
substantially reionized.

\section{Properties of the known $z > 6$ quasars and putative \ion{H}{2} regions}
\label{properties}

The Sloan Digital Sky Survey ({\it SDSS}) has discovered seven quasars
at $z \ge 6$ (Fan et al.  2001;2003;2004). These quasars are listed in
Table~\ref{tab1}. Column~2 lists the ultra-violet AB
magnitudes. Column~3 lists the source redshifts and column~4 lists the
redshift for the onset of the GP absorption (see below). Column~5
lists the difference between the host galaxy redshift (see below) and
the GP redshift, while column~6 gives the corresponding size ($R$) in
physical Mpc of the \ion{H}{2} region.  The typical size for
\ion{H}{2} regions around luminous quasars at $z\sim6$ is
$R\sim5$Mpc. In this paper we concentrate on investigating the
observability of \ion{H}{2} regions of this size.

\vspace{5mm}

\section{models of quasar \ion{H}{2} regions}
\label{model}

\begin{figure*}[t]
\epsscale{2.} \plotone{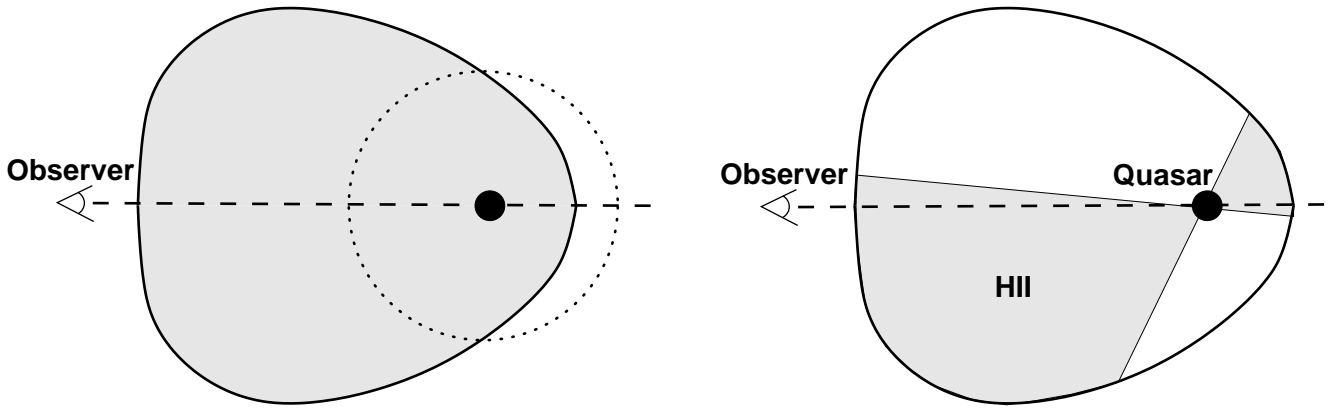}
\caption{\label{fig1} A schematic illustration of the observed \ion{H}{2} 
region geometry.  
The outer egg shaped boundary represents the shape of the surface of a
relativistically expanding spherical \ion{H}{2} region as seen by an
observer. The dotted circle in the left panel shows the intrinsic size
of the spherical \ion{H}{2} region at a time $t_{\rm age}$ that is $R/c$ prior to the
observation (see Wyithe \& Loeb~2004b). In the right panel we show an
example of a bi-conical
\ion{H}{2} region. Here the boundary of the 21cm emission is shaped by
the intersection of the equal observing-time surface with the
ionization cones of the quasar (shaded). The symmetry axis of a
conical \ion{H}{2} region may be misaligned with the line-of-sight
(dashed) as in the case illustrated on the right.}
\end{figure*}

Given a line-of-sight ionizing photon emission rate of $\dot{N}/4\pi$
photons per second per steradian and a uniform IGM, the line-of-sight
extent of a quasar's \ion{H}{2} region (from quasar to the {\em front}
of the region) observed at time $t$ is (White et al.~2003)
\begin{equation}
\label{ev}
R = 4\mbox{Mpc}\left[\frac{\dot{N}}{10^{57}}x_{\rm HI}^{-1}\frac{t_{\rm age}}{10^7\mbox{yr}}\left(\frac{1+z}{7.5}\right)^{-3}\right]^{1/3}.
\end{equation}
Here $x_{\rm HI}$ is the neutral fraction, and 
\begin{equation}
t_{\rm age} = t-R(t_{\rm age})/c
\end{equation}
is the age of the quasar corresponding to the time when photons that reach $R$ at time $t$ were emitted.
Using this relation between proper and retarded time we can compute
the observed shape of an \ion{H}{2} region given an intrinsic shape
(Wyithe \& Loeb~2004b; Yu~2004). Relativistically expanding \ion{H}{2}
regions appear beamed at the front and shortened at the back, an effect that is shown schematically in Figure~\ref{fig1}. While
the expansion velocity is not explicit in the formalism, it is
implicit in the specification of the observables $R$ and $\dot{N}$
through equation~(\ref{ev}).

In this paper we consider two examples for the intrinsic shape of the
\ion{H}{2} region. First we consider an intrinsically spherical \ion{H}{2}
region of radius 5Mpc. While a spherical \ion{H}{2} region has
traditionally been assumed, evidence for large obscuring dusty tori
surrounding quasars may imply that the geometry of the \ion{H}{2}
region is bi-conical. Indeed there is evidence for bi-conical
ionization and outflows in some AGN (e.g. Tadhunter \& Tsvetanov 1989;
Storchi-Bergmann, Wilson \& Baldwin 1992; Grimes, Kriss \& Espey
1999).  Therefore as our second example we consider a cone of length
5Mpc with an opening angle of $\pi/3$ radians. This cone is oriented
at various angles relative to the line-of-sight. We assume a neutral
fraction of unity ($x_{\rm HI}=1$), and $\dot{N}=10^{57}$ which is
typical of known luminous high redshift quasars.

\section{simulated observations of \ion{H}{2} regions}
\label{observations}

To simulate the observation of a quasar \ion{H}{2} region we embed the
simple \ion{H}{2} regions described above into a neutral IGM with
density fluctuations, add telescope noise to the resulting pixelized
image cube and then smooth the result to a given synthesized beam size
and frequency resolution in order to approximate the response of a synthesis
array to the \ion{H}{2} region.

Our simulations are generated at a resolution much higher than the
observations in both frequency and angle.  We start by adding
fluctuations in brightness temperature due to density fluctuations in
the IGM (at $z=6.5$). The fluctuations are added by generating a
realization of a pixelized density field described by an isotropic
power-spectrum, which on the scales of interest may be approximated by
the form $\langle\Delta T^2\rangle^{1/2}\propto \Delta\theta^{-0.3}$
(Furlanetto \& Briggs~2004). We smooth this field using a 10$^\prime$ top-hat beam and
a 0.5MHz top-hat frequency filter. The amplitudes of the pixelized density field are
then scaled so that the variance of the smoothed density field is 3mK
(Furlanetto \& Briggs~2004).  The \ion{H}{2} region is then embedded into
this density field. This is achieved by determining the {\em observed}
geometry of the \ion{H}{2} region, and then setting the brightness temperature contrast to
zero at all points inside the \ion{H}{2} region.

To approximate the response of a synthesis array we generate pixelized
noise cubes containing white noise in angular dimensions, and $1/f$
noise in the frequency domain. The pixelized cube is smoothed by the
synthesized beam (for which we assume a Gaussian) and a top-hat in
frequency with width corresponding to the resolution of the
observation. In this paper the beam size $\Delta\theta_{\rm beam}$ is
defined to be the variance of the Gaussian beam.  The brightness
temperature contrasts of the pixelized cube are then scaled so that
the resulting variance in the smoothed cube equals the value derived
from the radiometer equation~(\ref{noiseeqn}). The scaled, pixelized
noise cube is then added to the cube containing the emission
signature.

This procedure neglects the contribution of fore-grounds. A possible
scheme for removing foregrounds will be to exploit the spectral
flatness of the foregrounds and perform a continuum subtraction on the
spectra corresponding to each spatial pixel in the synthesized image
cube. One difficulty will be the frequency dependence of polarized
foregrounds, however this procedure should clean the cube of
foreground contribution, and leave fluctuations in the 21cm emission
that are detectable in space as well as frequency. Initial simulations
suggest that this will indeed be the case (C. Johnston et
al. in prep). The assumption that this cleaning can be performed
efficiently is implicit in the results of this paper where image slices
are shown. In addition to the image slices, we show line-of-sight
spectra. These spectra are made by averaging adjacent spectra in pixels
that fall within an aperture whose radius equals the size of the
synthesized beam. Thus the signal to noise in these spectra is higher
than expected from Figure~\ref{fig0} because the effective spatial
resolution is $\sim2\Delta\theta_{\rm beam}$.

As discussed in \S~\ref{21cmtel} there are three planned steps in the
development of electronically steerable low-frequency radio telescopes
capable of observing signatures of the epoch of reionization. Each
step will involve an order of magnitude increase in the collecting
area available, and allow higher resolution observations. In the
remainder of this section, we show examples of the observations
possible with each of these instruments.

\subsection{Observing \ion{H}{2} regions with the LFD}

\begin{figure*}[t]
\epsscale{2.} \plotone{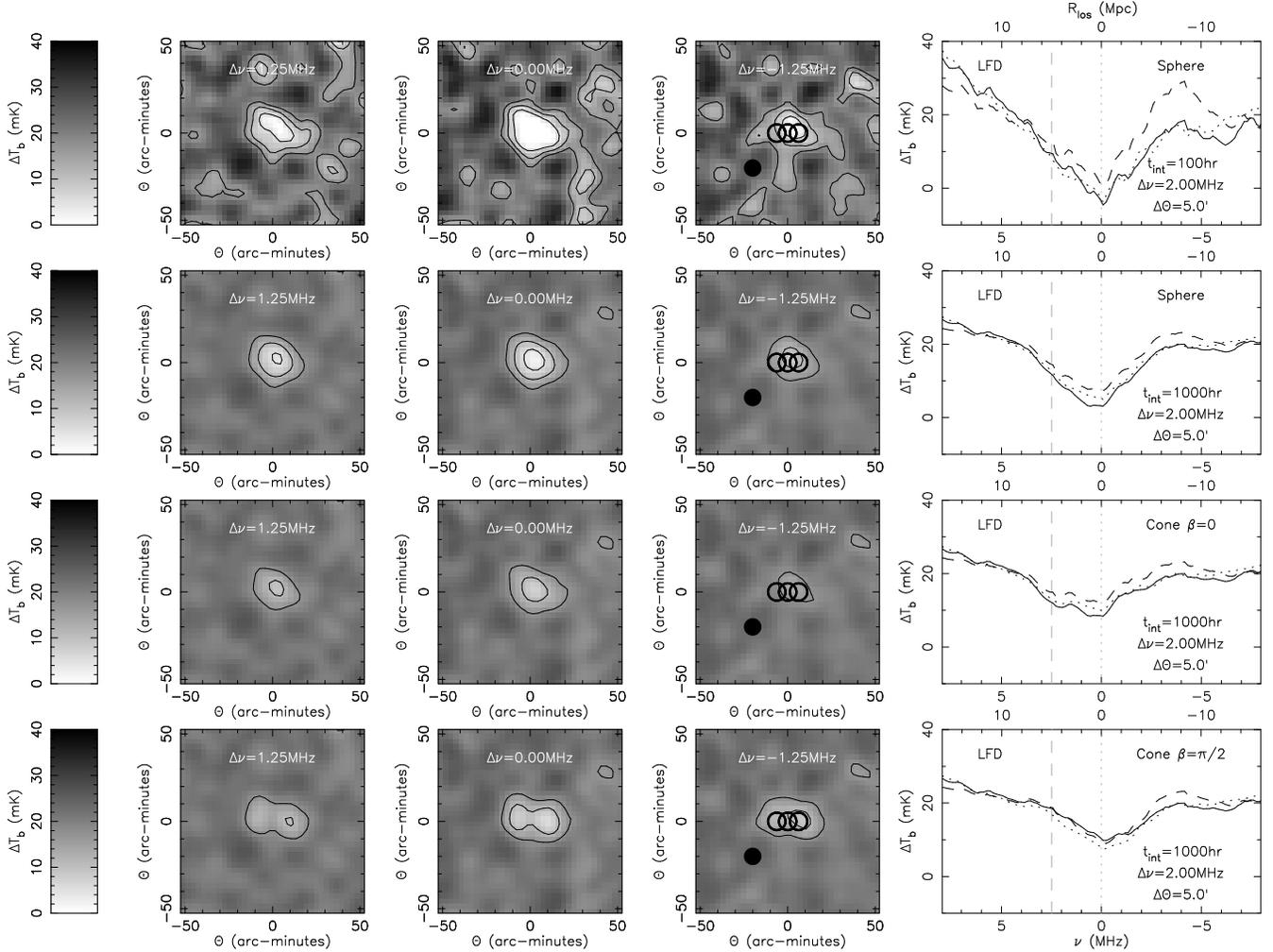}
\caption{\label{fig2} Sample maps of quasar HII regions observed
with the LFD around quasars at $z=6.5$. Four rows are shown. The first
and second rows correspond to a 100hr and 1000hr observation of an
intrinsically spherical \ion{H}{2} region. The observations were made
with a 5$^\prime$ Gaussian beam, top-hat smoothed at 2MHz.  The third and
fourth rows correspond to a 1000hr integration with a 5$^\prime$ Gaussian
beam, top-hat smoothed at 2MHz of an intrinsically bi-conical
\ion{H}{2} region with an opening angle of $\pi/3$ oriented along and
perpendicular to the line-of-sight respectively. In each case three
slices of the frequency cube are shown. The contours correspond to 5,
11 and 17mK. In addition, 3 smoothed spectra are shown in the
right-most panel. These are obtained by integrating over pixels inside
one beam radius from points along, and points $R_{\rm cone}/2$ left
and right of the line-of-sight. The beams at these points are plotted
on the third frequency slice. The dashed, solid and dotted spectra
correspond to the left, center and right lines-of-sight respectively,
while the dotted and dashed vertical lines show the redshifts of the
quasar and the front of the \ion{H}{2} region. The solid
black dot at (-20,-20) shows the beam size.}
\end{figure*}

\begin{figure*}[t]
\epsscale{2} \plotone{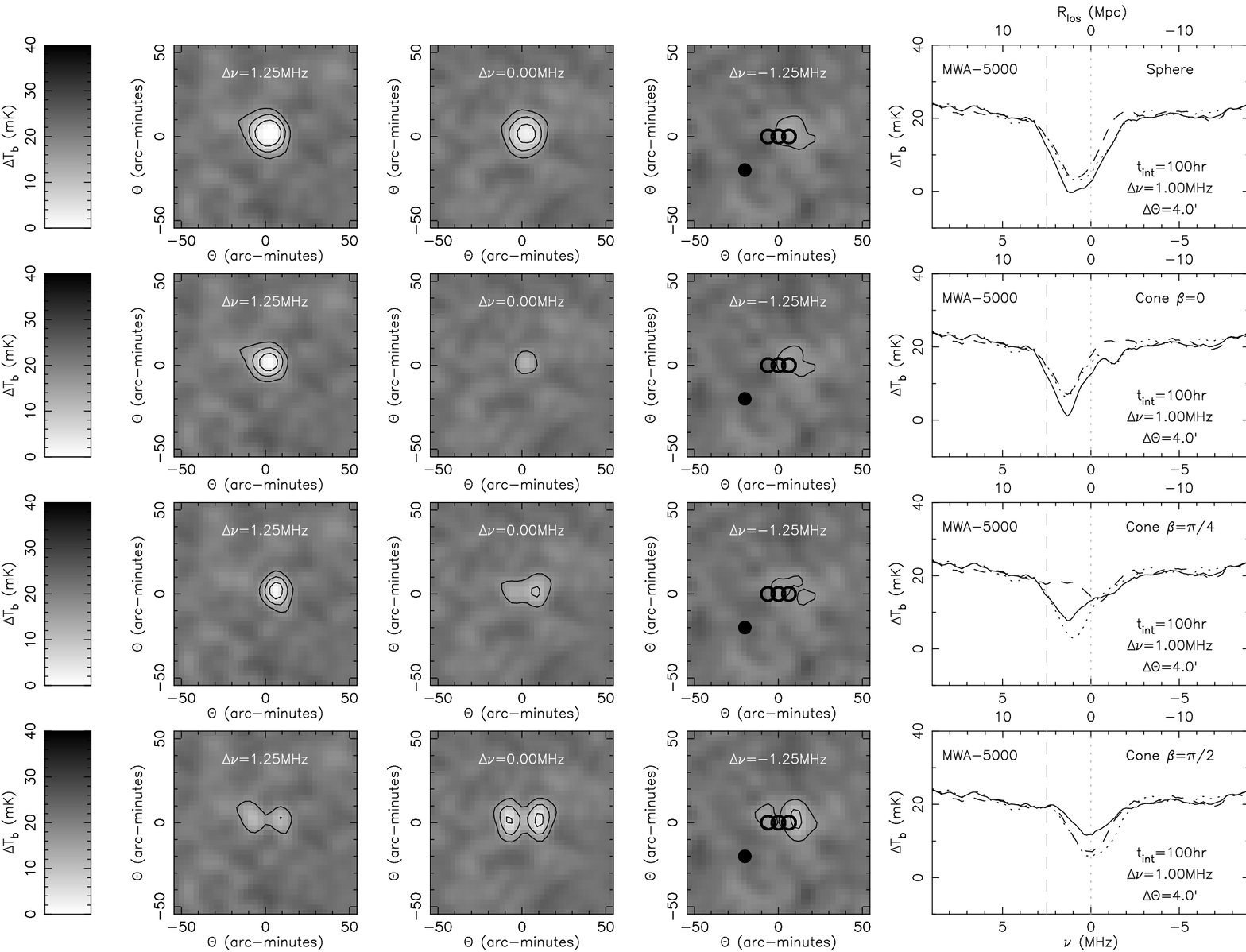}
\caption{\label{fig3} Sample maps of \ion{H}{2} regions 
around quasars at $z=6.5$ observed with the MWA-5000 with a 4$^\prime$ Gaussian beam and
top-hat smoothed at 1MHz in a 100hr integration. Four rows are shown. The first
row corresponds to observation of an intrinsically spherical \ion{H}{2}
region. The second, third and fourth rows correspond to observations
of an intrinsically bi-conical
\ion{H}{2} region with an opening angle of $\pi/3$ oriented along, at
$\pi/4$ to, and perpendicular to the line-of-sight respectively. In
each case three slices of the frequency cube are shown. The contours
correspond to 5, 11 and 17mK. In addition, 3 smoothed spectra are
shown in the right-most panel. These are obtained by integrating over
pixels inside one beam radius from points along, and points $R_{\rm
cone}/2$ left and right of the line-of-sight. The beams at these
points are plotted on the third frequency slice. The dashed, solid and
dotted spectra correspond to the left, center and right lines-of-sight
respectively, while the dotted and dashed vertical lines show the
redshifts of the quasar and the front of the \ion{H}{2} region.  The
solid black dot at (-20,-20) shows the beam size.}
\end{figure*}

Inspection of Figure~\ref{fig0} shows that the root-mean-square noise
falls in the shaded region for integrations longer than 100hrs and for
frequency smoothing greater than 1MHz. As a demonstration, the top row
of Figure~\ref{fig2} shows a simulated observation of a 5Mpc
intrinsically spherical \ion{H}{2} region. The integration time was
100 hours, the beam size was 5$^\prime$, and top-hat smoothing in
frequency space was applied at 2MHz. The location of the resulting
telescope noise is shown in Figure~\ref{fig0} by the label L100. The
left-most three panels show three different slices through the image
cube in frequency space.  The right-most panel shows three
spectra. These spectra were obtained by averaging pixels spatially
within a circular aperture having the scale of the beam. The three
spectra correspond to the three lines-of-sight shown by the set of
circles in the third frequency slice. These circles show the size of
the averaged regions for each spectrum. The left, center and right
lines-of-sight are shown by the dashed, solid and dotted lines. The
size of the beam is shown by the black dot at $\theta=(-20,-20)$.  The
presence of an \ion{H}{2} region could be inferred from this data, but
no information would be gleaned regarding its geometry. The relatively
small contrast with respect to the background might make serendipitous
discovery of \ion{H}{2} regions difficult. On the other hand, one
would have more confidence in this detection of an
\ion{H}{2} region if it were around a known high redshift quasar. In
particular if the \ion{H}{2} region is coincident with an already
known quasar, then the redshift of the edge of the \ion{H}{2} region
would have a known redshift from near-IR observations (dashed vertical
line in the right-hand panel).

The lower three rows show results from longer integrations of 1000
hours using a beam size of 5$^\prime$. The smoothing in frequency space was
applied at 2MHz.  The location of the resulting telescope noise is
shown in Figure~\ref{fig0} by the label L1000. Firstly the second row
repeats the observation of the sphere. The additional integration time
reduces the noise to well below the contrast of the \ion{H}{2}
region. As a result the \ion{H}{2} region is detected at high
significance, and the contrast is larger due to the smaller beam
size. However the size of the smoothing in frequency space 2MHz is
comparable to the size of the \ion{H}{2} region. Therefore the
sharpness of the edge of the \ion{H}{2} region cannot be determined.

The lower two rows show \ion{H}{2} regions with conical geometry. In
the 3rd row the cone is aligned with the line-of-sight, while in the
bottom row the cone is oriented perpendicular to the
line-of-sight. The resolution in frequency of these spectra is not
sufficiently small to show the absence of ionized hydrogen at the
redshift of the quasar where the cone is pointed along the
line-of-sight. However in the lower-row the smaller synthesized beam provides
sufficient resolution that the observation is able to see the two
individual lobes of the bi-conical \ion{H}{2} region. In addition, the averaged
spectra in the 1000hr integrations show evidence for asymmetry owing to
relativistic expansion in the cases of a spherical \ion{H}{2} region,
and a conical \ion{H}{2} region oriented along the line-of-sight
(second and third rows). Thus, under favorable circumstances the LFD
may already provide useful information regarding the existence, number
and even geometry of \ion{H}{2} regions.

\subsection{Observing \ion{H}{2} regions with the MWA-5000}

\begin{figure*}[t]
\epsscale{2.} \plotone{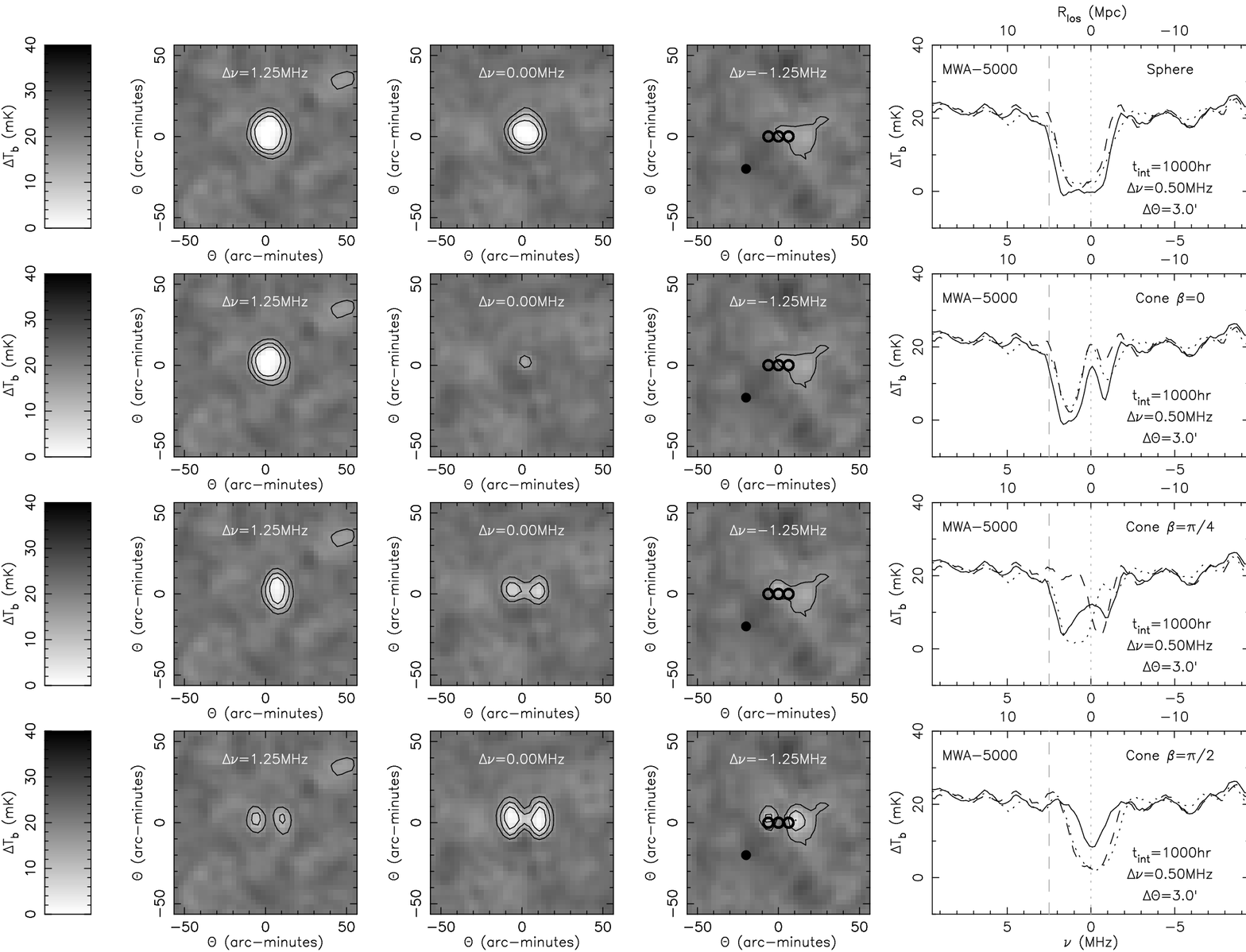}
\caption{\label{fig4} Sample maps of \ion{H}{2} regions observed 
around quasars at $z=6.5$ with the MWA-5000 with a 3$^\prime$ Gaussian beam
and top-hat smoothed at 0.5MHz in a 1000hr integration. Four rows are
shown. The first row corresponds to observation of an intrinsically
spherical \ion{H}{2} region. The second, third and fourth rows
correspond to observations of an intrinsically bi-conical \ion{H}{2}
region with an opening angle of $\pi/3$ oriented along, at $\pi/4$ to,
and perpendicular to the line-of-site respectively. In each case three
slices of the frequency cube are shown. The contours correspond to 5,
11 and 17mK. In addition, 3 smoothed spectra are shown in the
right-most panel. These are obtained by integrating over pixels inside
one beam radius from points along, and points $R_{\rm cone}/2$ left
and right of the line-of-sight. The beams at these points are plotted
on the third frequency slice. The dashed, solid and dotted spectra
correspond to the left, center and right lines-of-sight respectively,
while the dotted and dashed vertical lines show the redshifts of the
quasar and the front of the \ion{H}{2} region. The solid black dot at
(-20,-20) shows the beam size. }
\end{figure*}

\begin{figure*}[t]
\epsscale{2} \plotone{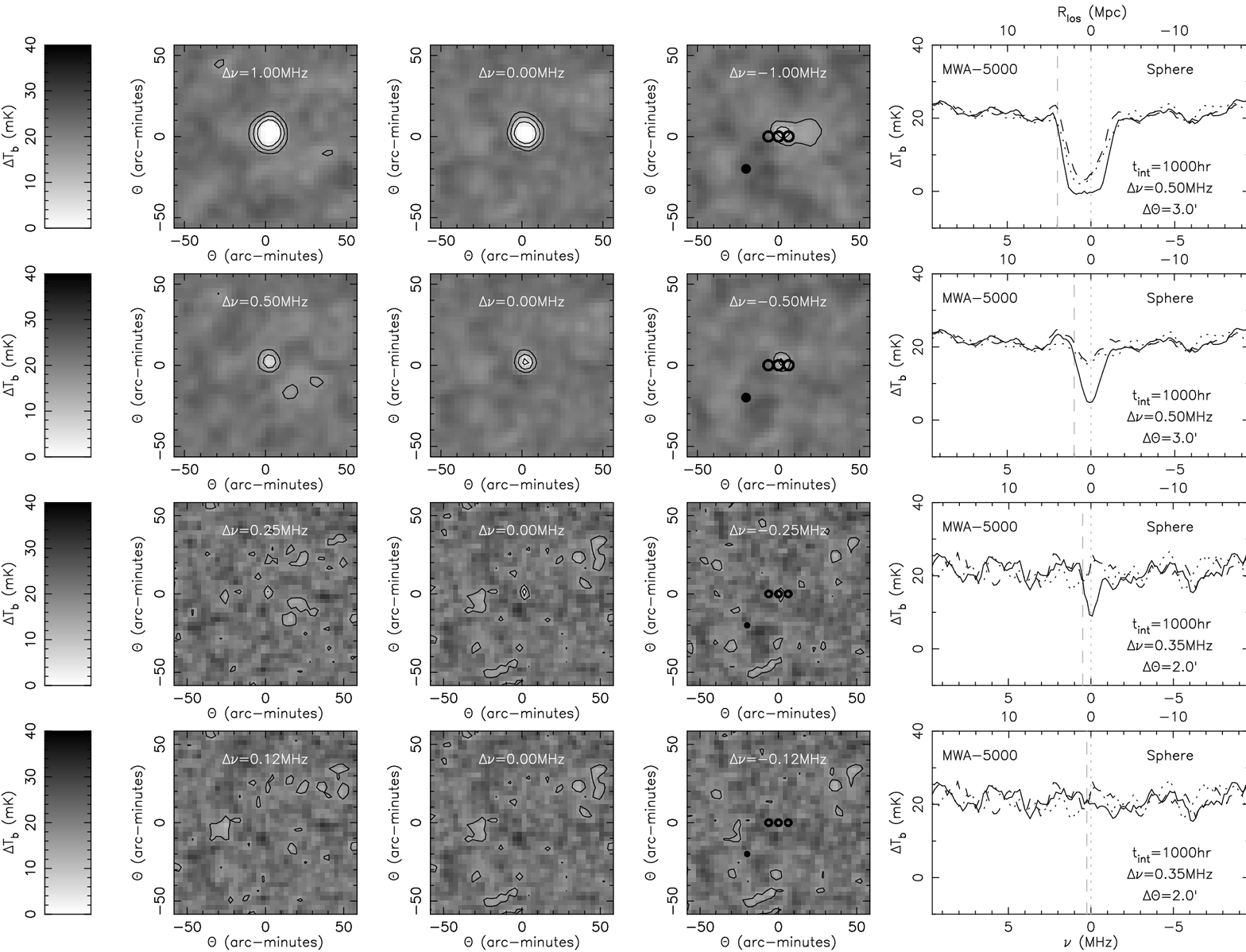}
\caption{\label{fig5} Sample maps of \ion{H}{2} regions observed
 around quasars at $z=6.5$ with the MWA-5000 in a 1000hr integration. Four
 rows are shown corresponding to intrinsically spherical \ion{H}{2}
 regions of size 4Mpc, 2Mpc, 1Mpc and 0.5Mpc. In the upper two rows
 observation was with a 3$^\prime$ Gaussian beam, top-hat smoothed at
 0.5MHz. In the lower two rows observation was with a 2$^\prime$ Gaussian beam
 top-hat smoothed at 0.35MHz. In each case three slices of the
 frequency cube are shown. The contours correspond to 5, 11 and
 17mK. In addition, 3 smoothed spectra are shown in the right-most
 panel. These are obtained by integrating over pixels inside one beam
 radius from points along, and points $R_{\rm cone}/2$ left and right
 of the line-of-sight. The beams at these points are plotted on the
 third frequency slice. The dashed, solid and dotted spectra correspond
 to the left, center and right lines-of-sight respectively,
while the dotted and dashed vertical lines show the redshifts of the
quasar and the front of the \ion{H}{2} region. The solid black dot at
 (-20,-20) shows the beam size.  }
\end{figure*}

While the LFD may be sufficient to detect \ion{H}{2} regions around
high redshift quasars under favorable circumstances, the MWA-5000 would be an
instrument with a collecting area sufficient to reach an instrumental
noise that is comparable to the cosmic web fluctuations at resolutions
significantly below the size of these \ion{H}{2} regions (see
Figure~\ref{fig0}). Here we show examples of observations of
\ion{H}{2} regions using MWA-5000 with integration times of 100 hours
(Figure~\ref{fig3}) and 1000 hours (Figure~\ref{fig4}). In each case
we show a spherical \ion{H}{2} region (upper rows), a conical
\ion{H}{2} region oriented along the line-of-sight (second rows), a
conical \ion{H}{2} region oriented at $\pi/4$ to the line-of-sight
(third rows), and a conical \ion{H}{2} region oriented perpendicular
to the line-of-sight (bottom rows). The additional collecting area of
the MWA-5000 allows observations to be made with a smaller synthesized
beam, and at higher spectral resolution. For the 100 hour observation
we adopt a beam size of 4$^\prime$, but decrease the smoothing in frequency
to 1MHz, while for the 1000 hour observation we reduce the beam size
to 3$^\prime$ and use a frequency smoothing of 0.5MHz. The resulting telescope
noise levels are plotted in Figure~\ref{fig0} using labels M100 and
M1000 respectively.

With these higher resolution observations the \ion{H}{2} regions are easily
detected. Moreover the signal to noise is sufficient that the \ion{H}{2}
regions could be discovered serendipitously in MWA-5000 fields. The geometry of
the individual \ion{H}{2} regions can be determined to some degree. For
example, in the 100hr integration the relativistic shortening of the back
of the bubble can be seen in the spherical case, and in the case where the
cone is directed along the line-of-sight. This can be seen either from the
frequency slices which show a lack of \ion{H}{2} region behind the bubble,
or from the averaged spectra which show that the minimum emission is at a
redshift blueward of the quasar.  The synthesized maps from the
100 hr observation reveal the conical geometry where the cone is oriented
away from the line-of-sight. Indeed where the cone is pointed at $\pi/4$ to
the line-of-sight the cone in front of the quasar appears larger due to
relativistic beaming. Inspection of the averaged spectra also reveals the
conical geometry. While the three lines-of-sight through the spherical
\ion{H}{2} region (top row) each find the same contrast, observations of a
conical \ion{H}{2} region oriented along the line-of-sight (second row)
show that the central spectrum (solid line) has the higher contrast because
observations from either side contain emission from neutral gas at the
redshift of the quasar. If the cone is oriented at $\pi/4$ to the
line-of-sight (third row) then the conical geometry and orientation, as
well as the relativistic expansion can be deduced from the relative depths
of the spectra, and the redshifts where the contrast is maximized. Finally,
if the cone is oriented perpendicular to the line-of-sight then the
observations either side of the line-of-sight reveal the same contrast at
the redshift of the quasar, while contrast along the line-of-sight is
reduced.

The situation is improved still further if the integration time is
increased to 1000 hours. The narrower frequency smoothing used in the longer
integration allows the edge of the \ion{H}{2} region to be better
defined. The rear edge of the \ion{H}{2} region occurs closer in redshift
to the quasar, showing the time delay effect owing to the relativistic
expansion. The morphologies described in the previous paragraph are more
prominent in this case. For example, in the second row, where the cone is
oriented along the line-of-sight, the high resolution and small frequency
bin result in the observation of a peak of emission at the redshift of the
quasar. This peak is the result of our idealized conical geometry. If the
host galaxy produced an inner spherical \ion{H}{2} region, this peak would
be less prominent in the observations.

The most luminous quasars discovered near the end of the reionization epoch
have putative \ion{H}{2} regions of 5Mpc.  These are the largest \ion{H}{2}
regions we expect to find, because quasars at higher redshift should be
less luminous. How small could the quasar \ion{H}{2} region be and still
remain detectable by the MWA-5000? Inspection of Figure~\ref{fig0} suggests that
in $\sim1000$ hours, regions of $\sim1$Mpc could just be detectable;
however this may be degraded by the presence of the fluctuations in the
IGM. In Figure~\ref{fig5} we show simulated observations of intrinsically
spherical \ion{H}{2} regions with the MWA-5000. The integration time was 1000
hours.  As before the left-most three panels show three different slices
through the image cube in frequency space.  The right-most panel shows
three spectra. These spectra were obtained by averaging pixels spatially
within a circular aperture having the scale of the beam. The three spectra
correspond to the three lines-of-sight shown by the three circles in the
third frequency slice (from left to right these spectra are shown by the
dashed, solid and dotted lines). The size of the beam is also shown by the
black dot at $\theta=(-20,-20)$.  The upper two rows show \ion{H}{2}
regions of $R=4$Mpc and $R=2$Mpc respectively for which the luminosity of
the source quasar was scaled down in proportion to $R^3$ relative to the
previous examples. The beam size was 3$^\prime$, and top-hat smoothing in frequency space
was applied at 0.5MHz, yielding similar spatial resolution in the
directions along and perpendicular to the line-of-sight. The \ion{H}{2}
regions of this size are readily detectable with MWA-5000 at $<1000$ hour
integration. Smaller \ion{H}{2} regions require a synthesized beam smaller
than 3$^\prime$. The lower two rows show \ion{H}{2} region of $R=1$Mpc and
$R=0.5$Mpc respectively. The luminosity of the source quasar was again
scaled down in proportion to $R^3$ relative to the previous examples. The
beam size was reduced to 2$^\prime$, and top-hat smoothing in frequency space was applied
at 0.35MHz. The $R\la2$Mpc \ion{H}{2} regions
will not be detectable using MWA-5000 with a 1000 hour integration.

\subsection{Observing \ion{H}{2} regions with the SKA}

\begin{figure*}[t]
\epsscale{2} \plotone{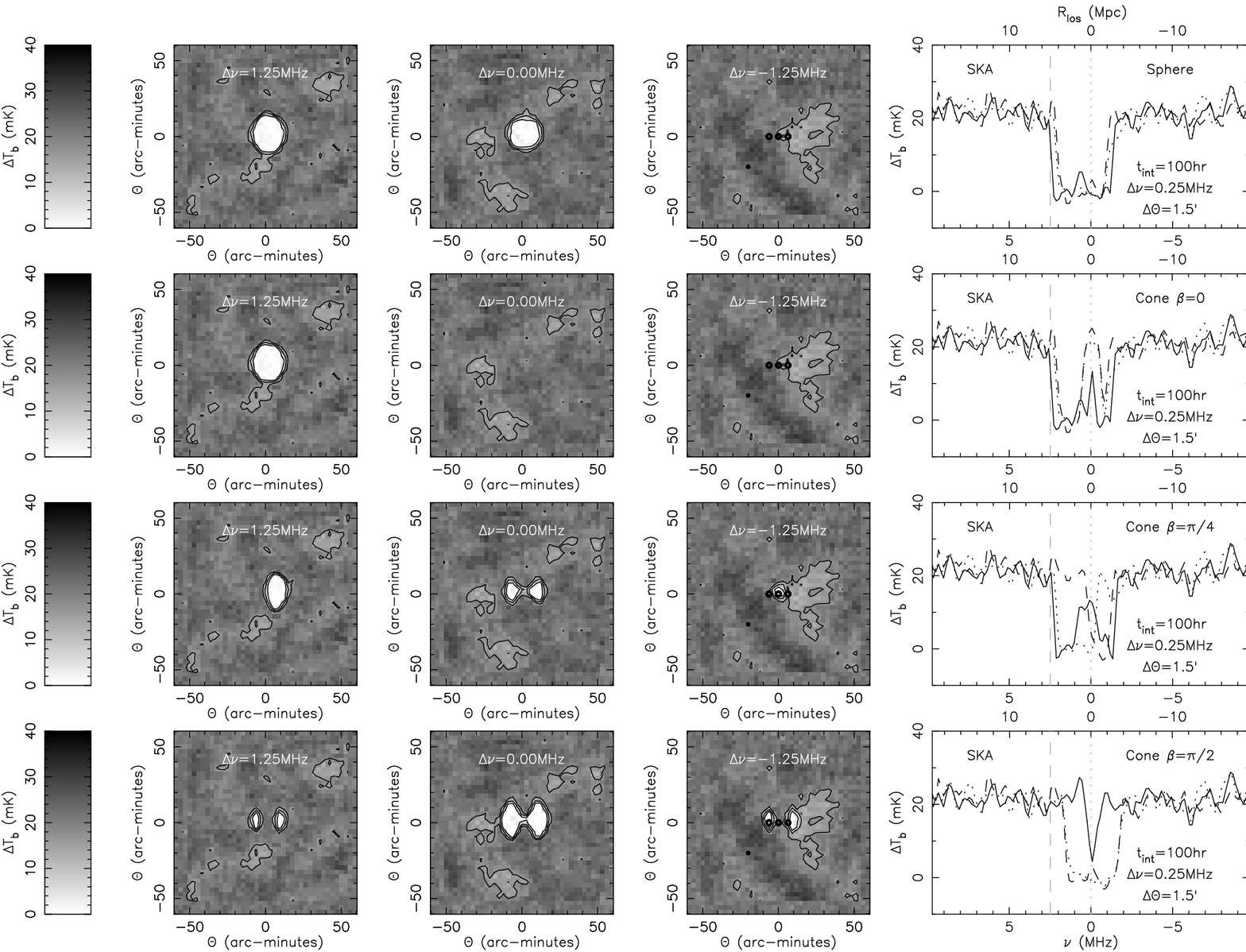}
\caption{\label{fig6} Sample maps of \ion{H}{2} regions observed
 around quasars at $z=6.5$ using the SKA with a 1.5$^\prime$ Gaussian
 beam and top-hat smoothed at 0.25MHz in a 100hr integration. Four
 rows are shown. The first row corresponds to observation of an
 intrinsically spherical \ion{H}{2} region. The second, third and
 fourth rows correspond to observations of an intrinsically bi-conical
 \ion{H}{2} region with an opening angle of $\pi/3$ oriented along, at
 $\pi/4$ to, and perpendicular to the line-of-sight. In each case
 three slices of the frequency cube are shown. The contours correspond
 to 5, 11 and 17mK. In addition, 3 smoothed spectra are shown in the
 right-most panel. These are obtained by integrating over pixels
 inside one beam radius from points along, and points $R_{\rm cone}/2$
 left and right of the line-of-sight. The beams at these points are
 plotted on the third frequency slice. The dashed, solid and dotted
 spectra correspond to the left, center and right lines-of-sight
 respectively, while the dotted and dashed vertical lines show the
 redshifts of the quasar and the front of the \ion{H}{2} region. The
 solid black dot at (-20,-20) shows the beam size.}
\end{figure*}

The LFD may be sufficient to detect \ion{H}{2} regions around high
redshift quasars and determine some aspects of their geometry. If the
integration time is extended to $\sim1000$ hours, then MWA-5000 will have
sufficient sensitivity and spectral resolution to enable serendipitous
discovery and detailed observation of \ion{H}{2} regions. However the
SKA would be an instrument with a collecting area sufficient to reach
an instrumental noise that is smaller than the cosmic web fluctuations
at resolutions much higher than the size of the \ion{H}{2} regions
(see Figure~\ref{fig0}), allowing detailed investigation of the
geometry of \ion{H}{2} regions. Here we show examples of observations of
\ion{H}{2} regions using SKA with integration times of 100 hours
(Figure~\ref{fig6}). Following the examples for the MWA-5000 we show a
spherical \ion{H}{2} region (upper row), a conical \ion{H}{2} region
oriented along the line-of-sight (second row), a conical \ion{H}{2}
region oriented at $\pi/4$ to the line-of-sight (third row), and a
conical \ion{H}{2} region oriented perpendicular to the line-of-sight
(bottom row). The additional collecting area of the SKA allows
observations to be made with a smaller synthesized beam, and at higher
spectral resolution than would be possible with the MWA-5000. We reduce the
beam size to 1.5$^\prime$ and use a top-hat frequency smoothing of width
0.25MHz. At $z=6.5$ this corresponds to comparable spatial resolution along
and perpendicular to the line-of-sight. The resulting telescope noise
is plotted in Figure~\ref{fig0} using the label S100.

As may be seen from the simulated maps in Figure~\ref{fig6} the SKA
allows observation of fine details of the \ion{H}{2} region geometry,
including all the morphologies described for the MWA-5000 observations, but
with higher signal-to-noise and at higher resolution. In particular
the high spectral resolution provides an excellent detection of the sharp
boundary of the \ion{H}{2} region, allowing its location in redshift
to be accurately determined. However Figure~\ref{fig6} demonstrates a
fundamental limitation in the observation of \ion{H}{2} regions which
is imposed by the density fluctuations in the IGM. These fluctuations
impose an irreducible noise in the synthesized maps for a telescope
with the sensitivity of SKA which increases in amplitude as the
resolution is increased. Consequently Figure~\ref{fig6} approximates
the best observation that can be made of an \ion{H}{2} region.

\section{what is the expected abundance of quasar \ion{H}{2} regions?}
\label{counts}

\begin{figure*}[t]
\epsscale{1.8} \plotone{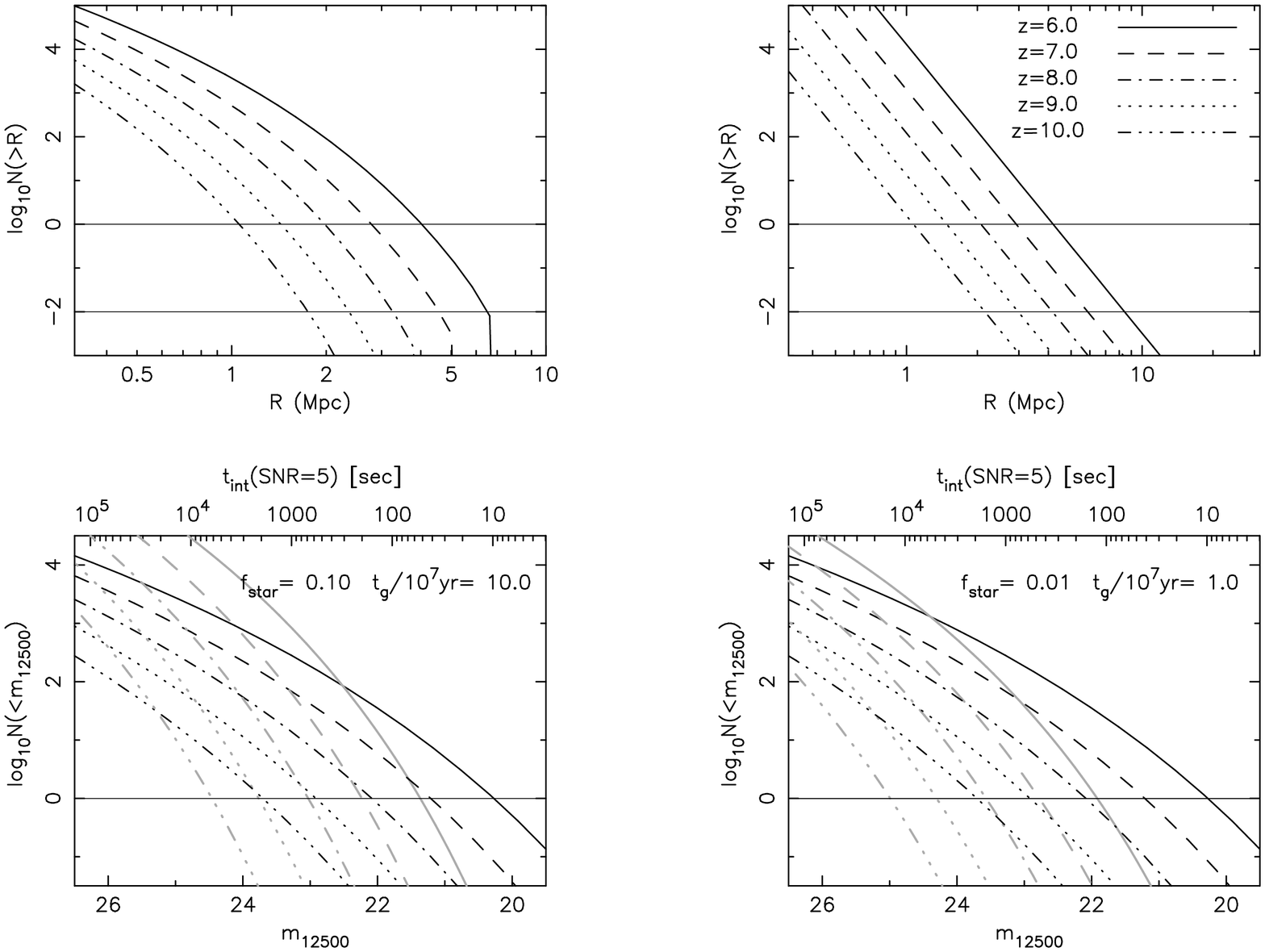}
\caption{\label{fig7} Number counts of high redshift galaxies and quasars
per 400 square degree - $16$MHz bandpass field . {\em Upper panels:} The
number of quasar \ion{H}{2} regions larger than $R$. We have converted from
luminosity to $R$ using equation~(\ref{ev}). In the left panel the number
counts were obtained from the model quasar luminosity function described in
Wyithe \& Loeb~(2003b). In the right-hand panel the counts were obtained by
extrapolating to high redshifts the luminosity function derived for
$z\sim6$ quasars (Fan et al.~2004), together with the redshift evolution of
quasar density (Fan et al.~2001b). Number counts are shown at redshifts
between 6 and 10. The horizontal line at unity guides the eye to the
largest \ion{H}{2} region around an active quasar in the field. A second
horizontal line at $10^{-2}$ corresponds to the largest fossil \ion{H}{2}
region (around a dormant quasar) in the field assuming a duty-cycle of
1\%. {\em Lower panels:} The dark lines show number counts of quasars
(using the model luminosity function of Wyithe \& Loeb~2003b) as a function
of apparent AB magnitude at 12500\AA. Also shown (grey lines) are the
number counts of starburst galaxies assuming $f_{\rm star}=0.1$,
$t_{g}=10^8$yr (left) and $f_{\rm star}=0.01$, $t_{g}=10^7$yr (right). The
luminosities were computed following the prescription in Loeb et
al. (2004), and the abundance from the Press-Schechter~(1974)
formalism. The upper axis shows an estimate of the integration time
required by the Gemini Telescope (computed using the Gemini exposure time
calculator). The horizontal line at unity guides the eye as to the brightest
sources in the field.}
\end{figure*}

The phased array design of planned low-frequency telescopes like the
MWA-5000 with a large number of stations each consisting of $4\times4$
crossed dipoles provides a primary beam with a diameter of $\sim20$
degrees. When combined with an expected bandpass of at least 16MHz, a
single observation will probe a very large volume. A single field is
therefore expected to contain significant numbers of luminous quasars,
even given their rarity at $z>6$ (Fan et al.~2004). Kohler, Gnedin,
Miralda-Escude \& Shaver~(2005) have investigated the statistics of
\ion{H}{2} regions using cosmological simulations combined with a
number density of quasars using an empirical extrapolation. They found
that an \ion{H}{2} region around a low-luminosity quasar should be present in
each synthesized beam. Here we concentrate on \ion{H}{2} regions
around luminous quasars whose existence is known at $z>6$ and find the
number of
\ion{H}{2} regions that we expect per primary rather than synthesized
beam as this is the quantity of more interest observationally.

We estimate the number counts of quasars per 400 square-degree field
in a band-pass of 16MHz as a function of B-band luminosity using two
methods. First, we generate number counts using the quasar luminosity
function model described in Wyithe \& Loeb~(2003b). Second we use an
empirical extrapolation of the luminosity function found for luminous
quasars at $z\sim6$ (Fan et al.~2004) combined with the evolution
found at $z\sim3-6$ (Fan et al.~2001b) and a spectral index of
$\alpha=-0.5$. We use equation~(\ref{ev}) to relate the luminosity to
an \ion{H}{2} region size. The top panels of Figure~\ref{fig7} show
the resulting number counts of \ion{H}{2} regions of physical radius
$R$ in each of these cases, and for redshifts between $z=6$ and
$z=10$. There should be approximately one 4Mpc \ion{H}{2} region per field at
$z\sim6$, which is consistent with our finding of $\sim$ 1 quasar of
the luminosity of the SDSS $z>6$ quasars. We estimate that there
should be $\sim1$ quasar \ion{H}{2} region of $R\sim3$Mpc at $z\sim7$,
$\sim1$ quasar \ion{H}{2} region of $R\sim2$Mpc at $z\sim8$, and
$\sim1$ quasar \ion{H}{2} region of $R\sim1$Mpc at $z\sim10$. 

However at these redshifts the recombination time is longer than the Hubble
time, implying that fossil \ion{H}{2} regions should remain in the IGM
after the quasars turn off. The number of such fossil \ion{H}{2} regions
depends on the duty-cycle. For quasar lifetimes of $\sim10^7$ years at
$z\sim6$ this implies a duty-cycle of $\sim0.01$. Similarly, the quasar
luminosity function model of Wyithe \& Loeb~(2003b) implies a duty-cycle of
$\sim0.005$. Fossil \ion{H}{2} regions may therefore be more common than
active quasars by two orders of magnitude. This results in $\sim100$
dormant quasar \ion{H}{2} regions of $R\sim3$Mpc at $z\sim7$, $\sim100$
\ion{H}{2} regions of $R\sim2$Mpc at $z\sim8$, and $\sim100$ \ion{H}{2}
regions of $R\sim1$Mpc at $z\sim10$. The increase in number results in an
increase in the size of the largest \ion{H}{2} region that might be found
at a given redshift. We might therefore expect that there should be $\sim1$
fossil quasar \ion{H}{2} region of $R\sim5$Mpc at $z\sim7$, $\sim1$ quasar
\ion{H}{2} region of $R\sim4$Mpc at $z\sim8$, and $\sim1$ quasar \ion{H}{2}
region of $R\sim2$Mpc at $z\sim10$.

Given the numbers of \ion{H}{2} regions that may be discovered in a
single MWA-5000 field, we suggest the possibility of using near-IR followup
of redshifted 21cm data as an efficient means to find high redshift
galaxies and quasars. In the lower panels of Figure~\ref{fig7} we
re-plot the number counts for high redshift quasars as a
function of the apparent AB magnitude at 12500 \AA (dark lines). 
Most regions will be fossil \ion{H}{2} cavities from
quasars that became dormant by the time of the observations. The
upper axes are labeled by the corresponding integration time for the
Gemini telescope\footnote{We used the integration time calculator
for NIRI on Gemini, see http://www.gemini.edu/sciops/instruments/instrumentITCIndex.html}. Note that even at these high redshifts the
bright quasars have $m_{12500}\sim20$ which is within reach of an 8m
class telescope in a short integration. Moreover the redshift of any
putative high redshift quasar would be approximately known (at the
center of the \ion{H}{2} region). It would therefore be practical to
take a snapshot survey of an MWA-5000 field to search for unknown
high-redshift quasars, even though the duty-cycle and therefore the
likelihood of finding an active quasar is low.

Of course galaxies will also be present at high redshift, will
generate their own \ion{H}{2} regions, and will be found within fossil
\ion{H}{2} regions of quasars. As a function of apparent magnitude we 
estimate the number counts of star-burst galaxies as
follows. Following Loeb, Barkana \& Hernquist~(2004) we find the mass
of a dark matter halo corresponding to the implied luminosity for a
fixed lifetime and star-formation efficiency. We have chosen sets of
$(f_{\rm star},t_{\rm g})=(0.1,10^8)$ and $(f_{\rm star},t_{\rm
g})=(0.01,10^7)$. These are combinations that produce less than 1
galaxy with $z<20$ per 5000 square degrees, as implied by the lack of
such galaxies at $z>6$ in the SDSS. These number counts are therefore
upper limits. The abundance of dark-matter halos multiplied by the
duty-cycle then gives the number-counts of star-burst galaxies per MWA-5000
field which are plotted (grey lines) in the lower-panels of
Figure~\ref{fig7} for comparison with the quasar number-counts. Bright
quasars are expected to be more common than bright galaxies as would
be expected from the results of the SDSS (Fan et al.~2004). The
largest \ion{H}{2} regions are therefore produced by quasars. For
example, at $z\sim7$ we expect quasars to be more common for objects
brighter than $m_{12500}\sim23.5$. We expect $\sim100$ \ion{H}{2}
regions per field from quasars of this luminosity. However these
\ion{H}{2} regions have sizes of only $R\sim1.3$Mpc which MWA-5000 will not
be able to observe.

\section{Science opportunities in the 21cm detection of intergalactic  \ion{H}{2} regions}
\label{science}

The results of the previous sections suggest some very interesting
possibilities for science outcomes based on redshifted 21cm observations of
\ion{H}{2} regions. These lie in the areas of the study of quasars and the
unified model, in addition to study of the reionization itself.

Firstly, the detection of contrast between an \ion{H}{2} region and a
warm IGM offers a direct measure of the neutral fraction at that
redshift. The serendipitous discovery of \ion{H}{2} regions that will
be feasible with the MWA-5000 (and possibly with the LFD) in addition to
the study of \ion{H}{2} regions around known high redshift quasars
will allow a determination of the neutral fraction as a function of
cosmic time. This will trace the evolution of neutral gas with
redshift, providing the same information as the previously discussed
global signature of reionization (Shaver et al.~1999). However the
{\em observed} evolution from neutral to ionized IGM has a minimum width of
$\Delta z\sim0.5$ due to a combination of cosmic variance and finite
light travel time (Wyithe \& Loeb~2004d). Since the global signature
must therefore extend over more than one band-pass rather than
occurring in a narrow step, it will be very difficult to calibrate. On
the other hand, the interior of \ion{H}{2} regions provide {\em built-in}
calibrators at each redshift where a measure of neutral fraction is
made. Measurement of the evolution of neutral fraction with time using
\ion{H}{2} regions will be complimentary to measurements of the
reionization history using the evolution in the power-spectrum of
redshifted 21cm fluctuations (Furlanetto, Hernquist \&
Zaldarriaga~2004). The latter will provide a richer data set than
observations of individual \ion{H}{2} regions. However both types of
measurement will be made using the same deep observations. They are
complimentary since they will be subject to different systematic
uncertainties.  

Secondly, the observation of luminous quasar \ion{H}{2} regions at a
resolution that allows some detection of the geometry offer rich
possibilities in AGN physics. For example, detection of the shape of
the \ion{H}{2} region will reveal the emission profile of the quasar. A
conical region would imply that a dusty torus is present in these
early quasars which obscures the ionizing emission. This phenomenon
would be analogous to collimated outflows and ionization cones observed around
low-redshift quasars. Detection of an asymmetry in the redshifts of
the front and back of the \ion{H}{2} region relative to the redshift
of the quasar imply a relativistic expansion, and yield an estimate of
the quasar lifetime (Wyithe \& Loeb~2004b). The quasar lifetime could
also be estimated from the duty-cycle. This quantity will be
determined from the fraction of \ion{H}{2} regions (with opening
angles that include the line-of-sight for conical \ion{H}{2} regions)
that host active quasars. A direct measurement of the duty-cycle
would yield important insight into the physics of the growth of early
super-massive black holes.

In Figure~\ref{fig7} we showed that early in reionization, before
clustering of sources becomes important in the generation of
\ion{H}{2} regions (Wyithe \& Loeb~2004c; Furlanetto et al.~2004)
\ion{H}{2} regions larger than 2Mpc are generated by quasars which,
while active, are detectable with Gemini in 200-300 seconds (with
little sensitivity to redshift).  Since there should be $\sim1$ such
active quasar per field at $z\sim8$, and $\sim100$ at $z\sim6$,
near-IR follow up of redshifted 21cm data may prove to be the most
efficient means of discovering very high redshift quasars in the
future, since only a hundred pointings per quasar will be required
rather than a blind search of $\sim400$ square degrees. \ion{H}{2}
regions larger than 2Mpc (those that will be found by MWA-5000) will be
dominated by quasars. However with the advent of SKA, \ion{H}{2}
regions smaller than 2Mpc in extent will be found.  The detection of
the lower luminosity quasars and luminous galaxies that generate these
\ion{H}{2} regions will require integration times of thousands of
seconds per pointing for an 8m class telescope, quickly rendering
searches of candidate \ion{H}{2} regions for galaxies and quasars
impractical. However the next generation of Extremely Large Telescopes
(ELT) will provide the collecting area needed to perform the required
integrations. For example the Giant Magellan Telescope would have an
effective aperture of greater than 20 square meters, and shorten the
required integration time by nearly an order of magnitude (with
additional improvement due to higher resolution imaging). This will
allow discovery of lower luminosity sources that generate the smaller
\ion{H}{2} regions early in the reionization. Thus a combination of
SKA and an ELT will allow us to follow the evolution of the topology
of reionization and the sources that powered that evolution on a
galaxy by galaxy basis.

\section{Conclusion}
\label{conclusion}

In this paper we have assessed the sensitivity of three generations of
planned low-frequency arrays, the LFD, MWA-5000 and SKA with respect to
observing quasar generated \ion{H}{2} regions. We find that an
instrument of the class of the planned LFD will be able to obtain good
signal to noise on \ion{H}{2} regions around the most luminous
quasars, and determine gross geometric properties in $\sim1000$
hours. The follow up MWA-5000 will be capable of mapping detailed geometry
of \ion{H}{2} regions, while SKA will be capable of detecting very
narrow spectral features and will determine the sharpness of the \ion{H}{2}
region boundary. An SKA will most likely be limited by irreducible
noise from fluctuations in the IGM itself. The MWA-5000 (and even the LFD
in favorable circumstances) will discover serendipitous \ion{H}{2}
regions. We have estimated the number of such \ion{H}{2} regions which
are generated by quasars based on the observed number counts of
quasars at $z>6$, and find that there should be several-tens
per field with radii larger than 4Mpc at $z\sim6-8$. At redshifts where the IGM
is predominantly neutral, large \ion{H}{2} regions will be generated
by quasars rather than by galaxies. Near-IR follow up of these
\ion{H}{2} regions with 8m class telescopes will be an efficient
method for location of high redshift luminous quasars.  Similarly, the
smaller \ion{H}{2} regions that will be identified by an SKA will
provide sites to search for galaxies and low-luminosity quasars using
the next generation of $\sim20-30$m class telescopes, which will be
sensitive enough to image normal galaxies and quasars in integrations
of less than a thousand seconds.  Detection of the massive host
galaxies of dormant quasars could help reveal the impact of quasars on
the evolution of galaxies. Thus, the study of \ion{H}{2} regions,
combined with targeted follow-up in the near-mid IR may provide the
most efficient method for finding high redshift galaxies and quasars,
and will in turn yield direct evidence for classes of sources
responsible for reionization.

\acknowledgements 
The authors would like to thank Miguel Morales for helpful comments
on the manuscript. This work was supported in part by NASA grant NAG
5-13292, and by NSF grants AST-0071019, AST-0204514 (for A.L.). JSBW
and DGB acknowledge the support of the Australian Research Council.

\end{document}